\documentclass[aps,pra,groupedaddress,twocolumn]{revtex4-2}

\usepackage{amsmath}
\usepackage{bm}
\usepackage{graphicx}

\begin{document}

%\title{Single-beam pump-probe scheme with elliptically polarized light for\\high-sensitivity atomic magnetometry in low-temperature\\alkali-metal vapor cells}

%\title{Level-crossing resonances of electromagnetically induced absorption in a single elliptically polarized light wave for atomic magnetometry applications}

%\title{Zero-field level-crossing resonances on open atomic transitions driving by elliptically polarized light wave in presence of buffer gas: Line narrowing and high contrast effects}

%\title{Level-crossing resonances on open atomic transitions in a single elliptically polarized light wave for atomic magnetometry applications}

%\title{Single-beam level-crossing resonances on open atomic transitions:\\Line narrowing, high contrast and applications to atomic magnetometry}

\title{Level-crossing resonances on open atomic transitions in a buffered Cs vapor cell:\\Linewidth narrowing, high contrast and applications to atomic magnetometry}


\author{D.V. Brazhnikov$^{1,2}$}
\email[Corresponding author: ]{brazhnikov@laser.nsc.ru}
\author{V.I. Vishnyakov$^1$}
\author{A.N. Goncharov$^{1,2,3}$}
\affiliation{$^1$Institute of Laser Physics SB RAS, 15B Lavrentyev Avenue, Novosibirsk 630090, Russia}
\affiliation{$^2$Novosibirsk State University, 1 Pirogov Street, Novosibirsk 630090, Russia}
\affiliation{$^3$Novosibirsk State Technical University, 20 Karl Marks Avenue, Novosibirsk 630073, Russia}

%\author{V.I. Vishnyakov}
%\affiliation{Institute of Laser Physics SB RAS, 15B Lavrentyev Avenue, Novosibirsk 630090, Russia}

\author{E. Alipieva$^{4}$}
\author{C. Andreeva$^{4,5}$}
\author{E. Taskova$^{4}$}
\affiliation{$^4$Institute of Electronics, Bulgarian Academy of Sciences, 72 Tsarigradsko Chaussee, Sofia 1784, Bulgaria}
\affiliation{$^5$Faculty of Physics, Sofia University "St. Kliment Ohridski", 5 James Bourchier Boulevard, Sofia 1164, Bulgaria}

%\author{A.N. Goncharov}
%\affiliation{Institute of Laser Physics SB RAS, 15B Lavrentyev Avenue, Novosibirsk 630090, Russia}
%\affiliation{Novosibirsk State University, 1 Pirogov Street, Novosibirsk 630090, Russia}
%\affiliation{Novosibirsk State Technical University, 20 Karl Marks Avenue, Novosibirsk 630073, Russia}

\date{\today}

\begin{abstract}
The ground-state Hanle effect (GSHE) in alkali-metal atomic vapors using a single circularly polarized light wave underlies one of the most robust and simplest techniques in modern atomic magnetometry. This effect causes a narrow (subnatural-width) resonance in the light wave intensity transmitted through an atomic vapor cell. Usually, GSHE-based sensors operate in the so-called spin-exchange-relaxation-free (SERF) regime to reduce the resonance linewidth. However, this regime requires a relatively high temperature of vapors ($\approx150^\circ$C or even higher), leading to large heat release and power consumption of the sensor head. Besides, without applying special measures, SERF regime significantly limits a dynamic range of magnetic field measurements. Here, we study a pump-probe scheme involving a single elliptically polarized light wave and a polarimetric detection technique. The wave is in resonance with two adjacent optical transitions in the cesium D$_1$ line ($\lambda$$\,=\,$$894.6$~nm) owing to their overlapping in presence of a buffer gas ($130$~Torr of neon). Using a small ($V\approx0.1$~cm$^3$) glass vapor cell, we demonstrate the possibility of observing subnatural-width resonances with a high contrast-to-width ratio (up to $\approx45$~\%/mG) under a low-temperature ($\approx60^\circ$C) regime of operation thanks to a strong light-induced circular dichroism in the medium. Basing on a $\Lambda$ scheme of atomic energy levels, we obtain explicit analytical expressions for the resonance's line shape. The model reveals a linewidth narrowing effect due to openness of the scheme of levels. This result is unusual for magneto-optical atomic spectroscopy because the openness is commonly considered as a undesirable effect degrading the resonance characteristics. By measuring noise voltage, we have figured out a $1.8$~pT/$\surd$Hz sensitivity of magnetic field measurements with a $60$~fT/$\surd$Hz sensitivity in the photon-shot-noise limit. The results, in general, contribute to the theory of GSHE resonances and can be also applied to development of a low-temperature high-sensitivity miniaturized magnetic field sensor with an extended dynamic range.
\end{abstract}

%\keywords{}

\maketitle

\section{Introduction}

Optically pumped (atomic) magnetometers is being a rapidly developed technology in modern magnetometry. Sensitivity of atomic magnetometers (AMs) has already reached that of superconducting quantum interference devices (SQUIDs) \cite{Kominis2003}. At the same time, AMs do not need for a cryogenic temperature and consume much less power compared to SQUIDs. AMs have already found relevant and exciting applications in medicine for magnetocardiography (MCG) \cite{MCG}, magnetoencephalography (MEG) \cite{MEG}, magnetomyography (MMG) \cite{MMG}, blood velocimetry \cite{blood} and other directions. They can be also used in biology for studying plant biomagnetism \cite{biomag} or observing nuclear magnetic resonances in biomolecules \cite{NMR}.  

One of the most robust and simplest techniques in atomic magnetometry is based on the ground-state Hanle effect (GSHE) linked with the zero-field level-crossing phenomenon \cite{Breit,GawlikReview,Budker2002}. In contrast to the excited-state Hanle effect studied by Wilhelm Hanle \cite{Hanle} and his contemporaries, the GSHE provides much narrower level-crossing resonances (LCRs), especially if a buffer gas or an antirelaxation coating of the cell walls is used. This feature immediately gave researchers an idea that the effect can be used for measuring very weak magnetic fields \cite{Alexandrov1967,DupontRoc1969}. First studied in cadmium vapors \cite{Lehmann1964}, the GSHE is nowadays used in quantum magnetometry mainly with He or alkali-metal vapors such as Rb, Cs and K.

A single circularly polarized light wave is usually used in the GSHE-based sensors both for pumping atoms and for probing their quantum state. To observe a nonlinear resonance in the light wave intensity transmitted through the vapor cell, a transverse magnetic field is slowly ($\lesssim200$~Hz) scanned around zero. This technique does not need for using an additional rf magnetic field as in some other types of AMs \cite{Bloom1962,AlexandrovReview}, simplifying the scheme and mitigating cross-talk problems between adjacent sensor heads in a multi-channel mode of operation. The simplicity of the scheme also provides a possibility for drastic miniaturization of the sensor, retaining high reliability and sensitivity of measurements \cite{Shah2007,Mhaskar2012}.

%All these advantages have already led to successful commercialization of the GSHE-based sensors \cite{QuSPIN}.

State-of-the-art Hanle sensors \cite{MCG,MEG,MMG,Shah2007,Mhaskar2012,Li2019} engage the spin-exchange relaxation-free (SERF) regime \cite{Allred2002} to reduce linewidth of the resonance. This helps to achieve the highest sensitivity ($\delta B$) owing to a simple relation: $\delta B\,\approx\,$FWHM/SNR with FWHM being the full width at half maximum of the resonance (in units of G) and SNR being the signal-to-noise ratio in a 1-Hz bandwidth. The SERF regime, however, requires high atomic density, i.e. an increased temperature of alkali-metal vapors which is usually much higher than $100^\circ$C. It also means relatively high power consumption and heat release of the sensor, especially in a multi-channel mode used in several medical applications.

As a further development of the GSHE-based sensing technique, it would be interesting to propose a technique that can provide a high sensitivity of measurements at much lower temperature of vapors ($T$$\lesssim60$$^\circ$C) in a small vapor cell ($V\ll1$~cm$^3$). Losing in the resonance linewidth without the SERF regime, a high sensitivity of the sensor can be obtained by increasing the SNR. Since it is proportional to the resonance contrast ($C$), there is a problem of increasing $C$ or, more precisely, a contrast-to-width ratio (CWR) of the zero-field LCR under a low-temperature regime.

At low temperature, the standard Hanle scheme with a single either circularly or linearly polarized light wave can demonstrate high contrast and/or high CWR of the resonances only for extended vapor cells ($V$$\gg$$1$~cm$^3$) \cite{Andreeva2002,Papoyan2002,Auzinsh2008,Gozzini2009,Marmugi2012,Ravi2017}, making it difficult to design a miniature sensor. Other types of AMs based on nonlinear Faraday rotation also require a relatively large vapor cell volume to achieve a sub-pT sensitivity at a low temperature of vapors \cite{Lucivero2014,Wilson2019}. Various pump-probe cw \cite{Klein2017,Gozzini2017,Brazhnikov2018,LeGal2019,Brazhnikov2019,Brazhnikov2021} or pulsed \cite{Nicolic2014,Lenci2019} light field configurations could help to overcome this problem. For instance, high-quality Hanle resonances have recently been observed in our pump-probe scheme with cesium vapors \cite{Brazhnikov2021}: $C$$\approx\,$$80\%$ was achieved at FWHM$\,\approx2$~mG, yielding large CWR of around $40\,\%/$mG. The light field was composed of two counterpropagating light waves of circular polarizations with opposite handedness ($\sigma^+\sigma^-$ configuration). Since a small ($5\times5\times5$~mm$^3$) cubic vapor cell was used, the scheme is attractive for developing a high-sensitivity miniaturized magnetic field sensor. A key advantage of the studied scheme consisted in a possibility of obtaining high-quality resonances under a low temperature of vapors ($T$$\,\le\,60$$^\circ$C).

%For instance, in \cite{Lenci2019}, the authors using a Ramsey-like technique in a cm-scale potassium vapor cell demonstrated contrast of LCR as high as 25\% at FWHM$\,\approx2$~mG, yielding relatively high CWR of around $12\,\%$/mG.

In spite of the fact that linearly and circularly polarized light beams are mainly used in atomic magnetometers, there are several pump-probe schemes with an elliptically polarized single beam that have been successfully applied to magnetic field sensing. For instance, an elliptically polarized off-resonant laser beam and a miniature glass $5$$\times$$5$$\times$$5$~mm$^3$ Rb vapor cell was used in \cite{Shah2009}. A zero-field LCR was observed in the rotation of polarization ellipse caused by a circular birefringence of the medium. To achieve a $7$~fT/$\surd$Hz sensitivity, the SERF regime was engaged. This method has further been developed to perform vector measurements under a near zero-field conditions \cite{Tang2021}. A polarization modulation Bell-Bloom-like scheme was studied in \cite{Ben-Kish2010}. The authors used a cm-scale ($2$$\times$$2$$\times$$5$~cm$^3$) Rb vapor cell to observe the coherent-population-trapping (CPT) resonances in the light wave transmission. The cell was heated to a relatively low temperature of $78^\circ$C. Basing on the provided contrast and linewidth measured values, the authors deduced a picotesla sensitivity of their technique. Another simple and efficient scheme with time-modulated ellipticity of light wave was proposed in a recent study using a $8$$\times$$8$$\times$$8$~mm$^3$ Cs vapor cell \cite{Petrenko2021}. The resonant circular component of the light was used for pumping the atoms into a so-called ``stretch'' Zeeman state, while the linear one was used to measure optical rotation induced by a circular birefringence of the medium owing to an off-resonant interaction with the pumped atoms. The cell was heated to a temperature of $90^\circ$C and a sensitivity of $15$~fT/$\surd$Hz was achieved. Resonant interaction of $^{87}$Rb atoms with a single light wave with an ellipticity that was just slightly differ from $45^\circ$ was studied in \cite{Ma2013}. Such a wave can be decomposed into a relatively strong pump wave and a weak probe wave with counter-rotating circular polarizations (here referred to as $\sigma^+$ and $\sigma^-$ components). The probe wave transmission was monitored separately by means of a polarimeter. A relatively low buffer gas pressure (25 Torr) in the cell allowed the authors to excite a single optical transition $F_g$$\,=$$2$$\to$$F_e$$\,=\,$$2$ in the D$_1$ line and to observe a LCR of electromagnetically-induced-transparency (EIT) type in a cm-scale low-temperature (62$^\circ$C) vapor cell. To reach a $25\%$ contrast of the resonance, however, a microwave field was required, meaning the need for using a bulky microwave cavity. Similar low-temperature technique, without a microwave field, was used then to demonstrate a three-axis AM with a sensitivity of $\sim10$~pT/$\surd$Hz in a cm-size Rb vapor cell \cite{Pradhan2019}.

Here, inspired by the results in the $\sigma^+\sigma^-$ configuration \cite{Brazhnikov2021}, we consider a relevant modification to that scheme. Namely, we propose to use a single resonant elliptically polarized wave instead of using two counterpropagating ones. Similarly to the scheme studied in \cite{Ma2013,Pradhan2019}, such a wave can be decomposed into the probe ($E_p$) and pump ($E_c$) waves. The probe wave transmission is monitored at a polarimeter. The experiments show that the same high-quality resonances can be observed in the proposed single-beam scheme with a low-temperature $5$$\times$$5$$\times$$5$~mm$^3$ Cs vapor cell as in the two-beam $\sigma^+\sigma^-$ configuration. Obviously, such a scheme with only one beam is much more attractive for creating a miniature sensor. Besides, the scheme provides an additional possibility for differential observation of the pump and probe wave transmission signals with the help of a balanced photodetector, suppressing some types of noise similarly to schemes with a polarization rotation \cite{Shah2007,Shah2009,Petrenko2021}.

It should be noted that, in contrast to a single-beam scheme proposed in \cite{Ma2013}, we observe the resonance of electromagnetically induced absorption (EIA) in the probe wave transmission instead of the EIT resonance. It occurs due to a relatively high buffer gas pressure in the cell where both optical transitions $F_g$$\,=$$4$$\to$$F_e$$\,=\,$$3$ and $F_g$$\,=$$4$$\to$$F_e$$\,=\,$$4$ in the D$_1$ line are excited by the light, allowing us to observe an extremely high contrast ($\gtrsim80\%$) of the resonance without the need for applying any microwave field as in \cite{Ma2013}. Preliminary measurements of the noise spectrum density reveals a 1.8 pT/$\surd$Hz sensitivity of measurements. However, an estimated shot-noise-limited sensitivity is $\approx60$~fT/$\surd$Hz that can be achieved after applying additional efforts for reduction of various noise sources in our scheme.

Finally, we develop a simplified theoretical model that allows obtaining explicit analytical solutions of the resonance's line shape in a wide range of pump wave intensity, i.e. beyond the perturbation theory approach. An openness-induced line-narrowing effect has been predicted by the theory and confirmed then by the experiments.

\section{Theory}\label{Sec:Theory}

Electromagnetically induced transparency or absorption effects are usually observed in atomic vapors as subnatural-width resonances either in a single-frequency magneto-optical (Hanle) configuration \cite{Pfleghaar1993,Dancheva2000} or in a two-frequency light field \cite{Boller1991,Akulshin1998}. Sign of the resonance (EIT or EIA) depends on different conditions of the experiment, such as a structure of involved energy levels in combination with light wave polarization and direction of magnetic field scan or presence of a residual magnetic field (in the Hanle configuration) \cite{Akulshin1998,Papoyan2002,Dimitrijevic2008,Yu2010,Noh2010,Grewal2015,Lazebnyi2015}, collisional depolarization of the excited state in buffered or antirelaxation-coated vapor cells \cite{Yu2010,Failache2003,JOSAB,Belfi2007} and others.

\begin{figure}[!t]
\centering
\includegraphics[width=0.95\linewidth]{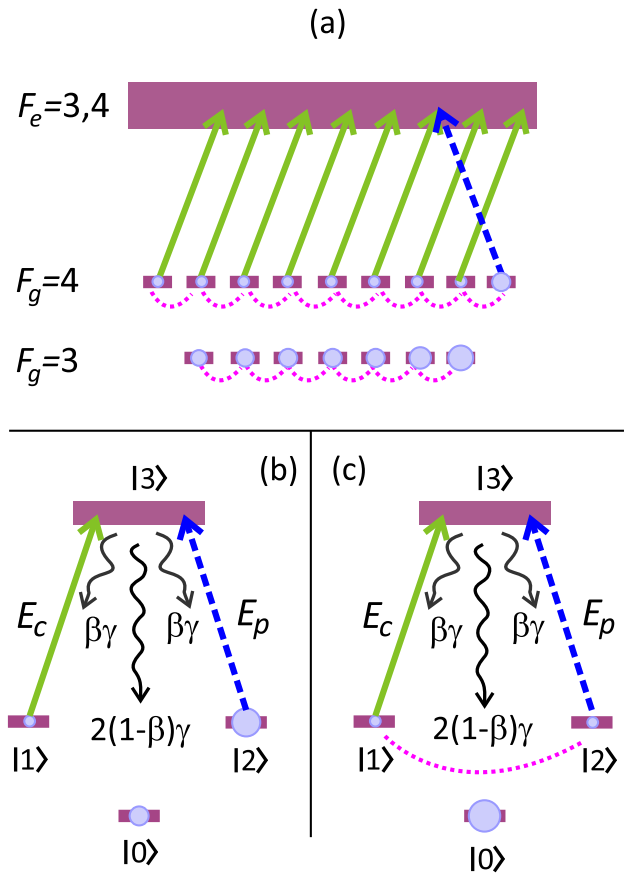}
\caption{\label{fig:1}(a) Scheme of energy levels in the D$_1$ line of Cs. Hyperfine structure in the upper state is not resolved due to buffer-gas collisional broadening. Zeeman sub-levels of the upper state are not shown. Solid green and dashed blue arrows stand for the pump and the probe waves, respectively (for simplicity, we show only one $\sigma^-$ transition for the probe light). Pink dotted lines in two ground states denote Zeeman coherences induced by the transverse magnetic field. Circles schematically reflect the sub-level populations. (b) Simplified three-level model of the atom (see details in the text).}
\end{figure}

Doppler line broadening or buffer-gas-collision broadening can also affect the resonance sign owing to a mutual influence of neighboring optical transitions in the D$_1$ or D$_2$ lines in alkali-metal atoms \cite{Akulshin1998,Papoyan2002,Grewal2015,Gozzini2009,Jadoon2022}). For instance, in \cite{Ma2013}, excitation of a single optical transition $F_g$$=$$2$$\to$$F_e$$=$$2$ in the D$_1$ line of $^{87}$Rb by an elliptically polarized light wave led to observation of the well-known EIT effect in the probe ($\sigma^-$) wave transmission as the result of destructive interference of the $\sigma^+$ and $\sigma^-$ optical transitions \cite{Arimondo1976}. In our scheme, on the contrary, we observe the EIA effect in the D$_1$ line of Cs under the similar conditions. The reason of such a behavior of the resonance can be explained by the significant overlapping of the $F_g$$=$$4$$\to$$F_e$$=$$3$ and $F_g$$=$$4$$\to$$F_e$$=$$4$ transitions in our case which are not spectroscopically resolved at 130 Torr of neon used as a buffer gas (the scheme of levels is shown in Fig. \ref{fig:1}a). The interference terms on these two transitions have opposite signs due to properties of the Clebsch-Gordan coefficients \cite{Varshalovich}, leading to significant suppression of the $\sigma^+$$-$$\sigma^-$ interference effects when the transitions are completely overlapped.

In our analysis, therefore, we consider two circularly polarized components of the light field to be independent, the probe $E_p(t,z)$ and the pump $E_c(t,z)$ waves, traveling along the $z$ axis:

\begin{equation}\label{eq:1}
    E(t,z) = E_c(z)\,e^{-i\omega t}  +E_p(z)\,e^{-i\omega t} + c.c.\,,
\end{equation}

\noindent where $E_{c,p}(z)$ are the real amplitudes slowly varying in space, $\omega$ is the optical frequency, ``{\it c.c.}'' means the complex conjugate terms. We assume that the buffer-gas broadening prevails over the Doppler broadening, so that atomic motion can be neglected. This explains the absence of $e^{\pm i k z}$ terms in (\ref{eq:1}) with $k$ being the absolute value of the wave vector.

In our experiments, field (\ref{eq:1}) is in resonance with the ``open'' $F_g$$=$$4$$\to$$F_e$$=$$3,\,4$ transitions (Fig. \ref{fig:1}a). Qualitative explanations for the reasons leading to observation of high-contrast EIA resonances in a probe wave transmission can be found in \cite{Brazhnikov2021}. Here, we consider a simplified three-level ($\Lambda$) scheme that, on the one hand, provides the similar explanation and, on the other side, allows us to derive explicit analytical solutions valid for a wide range of the pump wave intensity. In the scheme showed in Fig. \ref{fig:1}b,c, $|3\rangle$ is the excited state, while $|1\rangle$ and $|2\rangle$ stand for the ground state sub-levels of the same energy. These levels are in resonance with the light field. The trap state $|0\rangle$ is not in resonance with the field and serves just for collecting the atoms due to their spontaneous decay from the excited state. The dipole moments of two arms $|1\rangle$$\to$$|3\rangle$ and  $|2\rangle$$\to$$|3\rangle$ are assumed to be the same and equals $d_0$. It also means that the spontaneous decay rates of these arms are the same and can be denoted as $\beta\gamma$ with $2\gamma$ being the total spontaneous relaxation rate of the excited state ($2\gamma$$\,\approx\,$$2$$\pi$$\times4.56$~MHz for the Cs D$_1$ line), $\beta$ being the branching ratio that controls a degree of openness of the scheme: $0\le\beta\le1$ where $\beta$$\,=\,$$1$ corresponds to the case of a closed scheme. The ground-state relaxation is governed by the rate $\Gamma$ which is usually $\lesssim1$~kHz in the experiments with alkali-metal atoms in presence of a buffer gas (it is not shown in the figure for simplicity).

%Such a scheme is commonly used for theoretical study of various manifestations of the EIT \cite{Arimondo1996,Margalit2013,IEEE2018} and some types of the EIA \cite{LPL2014,EIAClock2019} effects.

%The EIA effect was first observed in a buffer-gas-free (evacuated) Rb vapor cell \cite{Akulshin1998,Akulshin1999} and was explained then as the result of spontaneous transfer of the Zeeman coherences \cite{EIA1999} and populations \cite{Goren2004} in ``closed'' two-level systems with degenerate energy levels on a so-called ``bright'' type transitions \cite{Lazebnyi2015}: $F_g$$=$$F$$\to$$F_e$$=$$F+1$ with $F$ being a nonzero integer number. However, under suitable conditions, the EIA effect can be also observed on ``dark'' type transitions, $F_g$$=$$F$$\to$$F_e$$=$$F,\,F$$-$$1$, where the EIT is usually observed. 

We use a standard quantum mechanics formalism of the density matrix $\hat{\rho}(z,t)$ to describe the atom-field interaction (for instance, see \cite{Blum}). The corresponding master equation has the Lindblad form:

\begin{equation}\label{eq:2}
\frac{\partial}{\partial t} \hat{\rho} = -\frac{i}{\hbar}\Bigl[\bigl(\hat{H}_0+\hat{V}_{b}+\hat{V}_{e}\bigr),\hat{\rho}\Bigr]+\hat{\cal R} \{\hat{\rho}\}\,,
\end{equation}

\noindent where the square brackets $\bigl[\ldots,\ldots\bigr]$ stand for the commutation operation of two matrices, $\hat{H}_0$ is a part of the total Hamiltonian for a free atom, $\hat{V}_b$ and $\hat{V}_e$ describe the interaction between the atoms and the magnetic and the light fields, respectively (in the electric-dipole approximation). The linear functional $\hat{\cal R}$ is responsible for the relaxation processes in the atom. The details can be found in Appendix.

%The light-induced recoil effect as well as the radiation trapping effect is not relevant under the considered conditions and can be neglected.

We assume the probe wave to be weak enough, so that optical properties of the medium do not depend on $E_p$. It means the linear approximation with respect to the probe field. In the experiments, a photodetector registers the wave intensity $I_{p,c}$$\,=\,$$(c/2\pi)E_{p,c}^2$ with $c$ being the speed of light. We first analyse the LCR in the pump wave intensity, which obeys the following equation (see Appendix): 

\begin{equation}\label{eq:3}
\frac{d I_c}{d z} = -\alpha_c I_c\,.
\end{equation}

%Its amplitude varies in space according to the reduced wave equation:

%\begin{equation}\label{eq:WaveEq}
%\frac{d E_p}{d z} = 2\pi i k P_p\,,
%\end{equation}

%\noindent where $P_p$ is the complex slowly varying amplitude of the atomic vapor polarization on the $|2\rangle$$\to$$|3\rangle$ transition. 

\noindent Here, the pump wave absorption index is

\begin{equation}\label{eq:4}
\alpha_c \approx \frac{3\gamma\beta\lambda^2 n_a}{4\pi\gamma_{eg}}\rho_{11}\,,
\end{equation}

\noindent where $\rho_{11}$ is the population of $|1\rangle$ sub-level, $\lambda$ is the light wavelength ($\approx\,$894.5~nm for the Cs D$_1$ line), $n_a$ is the atomic number density, which strongly depends on a temperature of the vapors, $\gamma_{eg}$ is the relaxation rate of the optical coherences in the $\Lambda$-scheme, determining the linewidth of the optical transitions: $\gamma_{eg}$$\,=\,$$\Gamma$$+$$\gamma$$+$$\gamma_c$ with $\gamma_c$ being the rate of dephasing collisions between the cesium atoms and the buffer-gas atoms. In our experiments, the collisional line broadening significantly prevails over the other impacts, i.e. $\gamma_c$$\,\gg\,$$\gamma,\,\Gamma$ meaning $\gamma_{eg}$$\,\approx\,$$\gamma_c$.

%Eq. (\ref{eq:AbsIndexP}) has been figured out using several approximations. In particular, we neglected the interference effect between the $E_c$ and $E_p$ waves as it is discussed above. Therefore, there is no a contribution proportional to the ground-state Zeeman coherence $\rho_{12}$. Besides, 

In the linear regime on the probe wave, the absorption index $\alpha_c$ depends only on the pump wave intensity $I_c$ that in turn slowly varies along the z axis due to absorption in the cell. However, as we will see, the pump wave experiences small absorption in the cell, i.e. the medium is optically thin for this wave (in a considered temperature of the cell). Therefore, $\alpha_c$ depends on $I_c$ as a constant parameter. This approximation allows us to obtain a simple solution in the form of the well-known Beer–Lambert–Bouguer law:

\begin{equation}\label{eq:5}
I_c(z) =I_{c0}\,e^{-\alpha_c z}\,.
\end{equation}

\noindent with $I_{c0}$ being the pump wave intensity at the entrance of the cell. For theoretical analysis, it is convenient to define a dimensionless coefficient of transmission following the expression: 

\begin{equation}\label{eq:6}
\eta_c = \frac{I_c(z=L_{cell})}{I_{c0}}=e^{-\alpha_c L_{cell}}\,,
\end{equation}

\noindent where $L_{cell}$ is the cell length.

Substituting $\rho_{11}$ from Appendix into (\ref{eq:4}), we get a compact solution:

\begin{equation}\label{eq:7}
\alpha_c\approx \frac{\alpha_0\,(1+\xi+4\Omega^2\tau^2)}{(1+\xi)\bigl[1+(2-\beta)\xi\bigr]+4\bigl[1+(1-\beta)\xi\bigr]\Omega^2\tau^2}.
\end{equation}

%\begin{equation}\label{eq:AbsIndexC2}
%\alpha_p \approx \alpha_0\, \frac{(1+\xi)(1+2\xi)+4\Omega^2\tau^2}{(1+\xi)\bigl[1+(2-\beta)\xi\bigr]+4\bigl[1+(1-\beta)\xi\bigr]\,\Omega^2\tau^2}\,.
%\end{equation}

\noindent Here, $\Omega$$\,=\,$$g B_x$ is the Larmor frequency where $g$ is the gyromagnetic ratio ($\approx2\pi\times370$~Hz/mG for the Cs ground state), $B_x$ is the transverse magnetic field strength, $\tau$$\,=\,$$\Gamma^{-1}$ is the relaxation time of the ground state, $\xi$ is the modified saturation parameter:

%$\mu_B$$=$$927.4\times10^{-23}$~erg/G

\begin{equation}\label{eq:8}
\xi = R_c^2\tau/\gamma_{eg} = \frac{3\beta\gamma\tau\lambda^3}{16\pi^2\hbar c \gamma_{eg}} I_c\,,
\end{equation}

\noindent which is nothing but the optical pumping rate, $R_c^2/\gamma_{eg}$, multiplied by the time of coherent interaction, $\tau$, where $R_c$$\,=\,$$d_0 E_c/\hbar$ is the Rabi frequency for the pump field. In (\ref{eq:8}), we took into account the relation: $\beta\gamma$$\,=\,$$4k^3d_0^2/3\hbar$.

Coefficient $\alpha_0$ in (\ref{eq:7}) is the absorption index at the center of the resonance curve ($\Omega$$\,=\,$$0$) in the low intensity limit ($\xi$$\,\ll\,$$1$):

\begin{equation}\label{eq:9}
\alpha_0 = \frac{3\beta\gamma\lambda^2 n_a \rho_0}{4\pi\gamma_{eg}}\,,
\end{equation}

\noindent with $\rho_0$ being the initial population of a sub-level in the ground state. In our model, we take $\rho_0$$\,=\,$$1/3$ for each of the $|0\rangle$, $|1\rangle$ and $|2\rangle$ sub-levels.

The other approximations of the theory that should be mentioned are the following. We neglect the effect of optical line splitting that can be observed when $\Omega$$\,\sim\,$$\gamma_{eg}$, because under our experimental conditions the resonance width is much less than $\gamma_{eg}$, so that we can focus on the range of $\Omega$$\,\ll\,$$\gamma_{eg}$. We can also consider the condition $R_c^2\ll\gamma_{eg}\gamma$ to be fulfilled under a reasonable light wave intensity. The condition $\gamma\tau$$\,\gg\,$$1$ is also satisfied in presence of a buffer gas when the time of coherent interaction between the light and the atoms are significantly increased in comparison with the case of a vacuumated vapor cell. Finally, since we neglected any interference effects between the probe and the pump waves, the subnatural-width-resonance splitting effect \cite{Alipieva2004,Margalit2013,Zhang2019} does not occur.

\begin{figure}[!t]
\centering
\includegraphics[width=\linewidth]{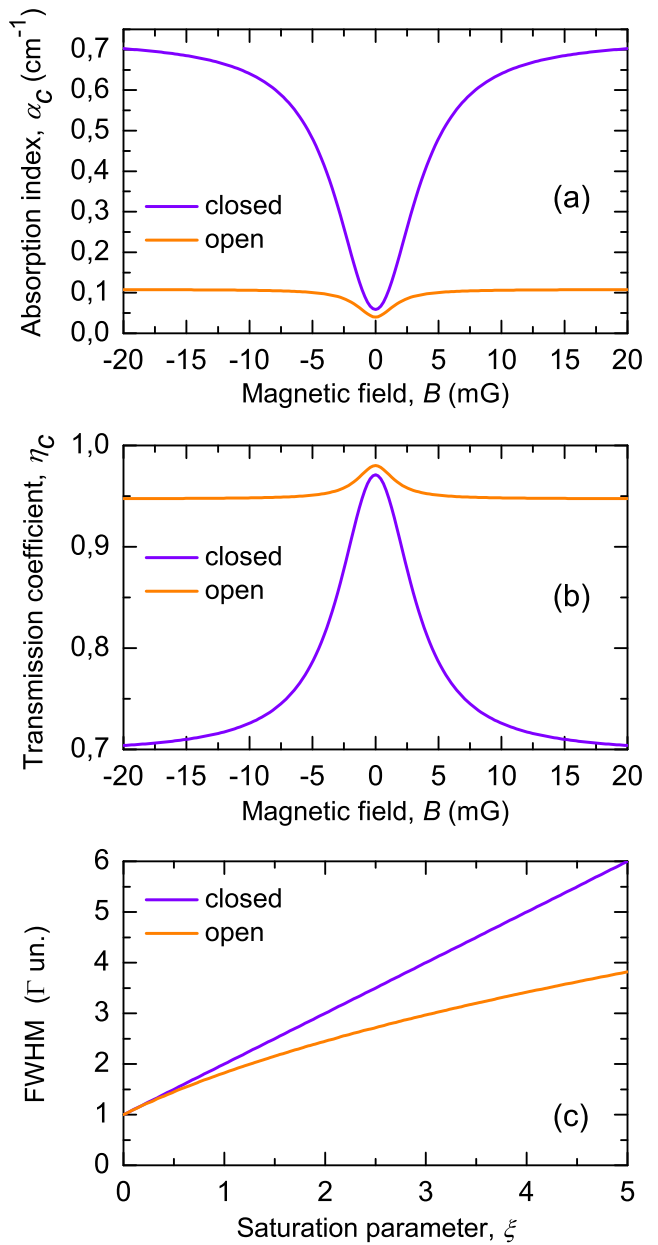}
\caption{\label{fig:2}Calculated zero-field level-crossing resonance of EIT in the pump-wave absorption index (a) and transmission coefficient (b) in closed ($\beta\,$$=$$\,1$) and open ($\beta\,$$=$$\,0.5$) system of energy levels. (c) Full width at half maximum of the resonance as a function of light wave intensity (dimensionless saturation parameter). Parameters of the calculation: $2\gamma$$\,=\,$$2\pi$$\times$$4.56$~MHz, $\Gamma$$\,=\,$$10^{-4}\gamma$, $\gamma_c$$\,=\,$$2\pi$$\times$$2$~GHz, $n_a$$\,=\,$$10^{12}$~cm$^{-3}$, $\lambda$$\,=\,$$894.4$~nm, $L_{cell}$$\,=\,$$0.5$~cm. In (a) and (b), Rabi frequency $R_c$$\,=\,$$\gamma$.}
\end{figure}

Fig. \ref{fig:2}a shows the behavior of $\alpha_c(\Omega)$ calculated for typical experimental conditions at $\beta$$\,=\,$$1$ and $\beta$$\,=\,$$0.5$. Fig. \ref{fig:2}b demonstrates the corresponding transmission coefficient $\eta_c(\Omega)$. The EIT resonance is observed with significantly suppressed height in the case of an open system of levels ($\beta$$\,=\,$$0.5$). The pump wave absorption is small both at $\Omega$$\,=\,$$0$ and $\Omega$$\,\ne\,$$0$, especially if the system is open. It makes our approximation concerning the low pump wave absorption in the cell to be valid.

The function $\alpha_c(\Omega)$ has the Lorentzian-like shape. The low pump wave absorption means that $\eta_c(\Omega)$ is also described by the same line shape owing to the approximation:

\begin{equation}\label{eq:10}
    \eta_c(\Omega)\approx 1-\alpha_c(\Omega)L_{cell}.
\end{equation}

\noindent The resonance linewidth (FWHM) in $\alpha_c(\Omega)$ and $\eta_c(\Omega)$ can be easily figured out from (\ref{eq:7}):

\begin{equation}\label{eq:11}
    \Delta_c \approx \Gamma \sqrt{(1+\xi)\,\Biggl[1+\frac{\xi}{1+(1-\beta)\xi}\Biggr]}\,.
\end{equation}

\noindent The effect of openness induced line narrowing readily follows the latter expression. Indeed, if the system of levels is closed ($\beta$$\,=\,$$1$), then ones get a linear behavior: $\Delta_c$$\,\approx\,$$\Gamma$$\,+\,$$R_c^2/\gamma_{eg}$$\,\propto\,$$I_c$. Such a dependence is well known in the theory of GSHE \cite{Allred2002,Cohen1970} as well as in the theory of two-frequency EIT/CPT effects \cite{Vanier1998,Lee2003} for homogeneously broadened closed transitions. In particular, the latest analytical results for a real (degenerate) structure of energy levels can be found in \cite{LeGal2022}. Otherwise, in open systems ($\beta\,$$<$$\,1$), the linewidth demonstrates a square-root-like behavior, $\Delta_c$$\,\propto\,$$\surd I_c$, which is clearer at $\xi$$\,\gg\,$$1$. The similar behavior has been recently noted in \cite{Pollock2022} for two-frequency CPT resonances in a buffered vapor cell. We demonstrate this narrowing effect in Fig. \ref{fig:2}c. To the best of our knowledge, such an openness-induced line-narrowing effect has not yet been emphasized in the literature for the magneto-optical (GSHE) resonances as well as it has not been drawn a special attention as a separate line-narrowing effect in the two-frequency CPT experiments.

%It is probably due to the fact that the GSHE resonances were thoroughly studied in closed schemes \cite{XX}, in open schemes with Doppler line broadening \cite{Gozzini2017,Grewal2015,Renzoni1997,Ram2010} or under the conditions when ground-state hyperfine structure is not resolved \cite{Allred2002,Gozzini2017}.

%The same dependency also appears in the theory of two-frequency CPT resonances under strong buffer-gas collisions \cite{Vanier1998} when both ground-state hyperfine levels contribute to the resonance creation.

The resonance contrast ($C$) can be defined as $C$$\,=\,$$[\eta_c(0)$$-$$\eta_c(\infty)]$$/$$\eta_c(0)$ where $[\eta_c(0)$$-$$\eta_c(\infty)]$ is the resonance height, $\eta_c(\infty)$ means the background, i.e. the light transmission at $\Omega$$\,\gg\,$$\Delta_c$. As seen from Fig.\ref{fig:2}b, in the standard Hanle configuration with only one circularly polarized wave for pumping and probing, the resonance contrast is quite low ($<\,$5\%) under a temperature of $\le\,$60$^\circ$C. The large difference between the resonance heights in closed and open systems (violet and orange curves in Fig.\ref{fig:2}a,b) can be easily explained. Indeed, at $\Omega$$\,=\,$$0$ (Fig.\ref{fig:1}b), the pump wave experiences small absorption in the cell owing to optical pumping most of the atoms to the non-interacting state $|2\rangle$ (in the closed scheme) or to both non-interacting states $|2\rangle$ and $|0\rangle$ (in the open scheme). The probe light is assumed to have a so weak strength that it does not noticeably affect the sub-level populations. Then, if the system of levels is closed and $\Omega$$\,\gg\,$$\Delta_c$, the two sub-levels $|1\rangle$ and $|2\rangle$ is strongly mixed, i.e. the so-called Zeeman coherence is created (pink dotted line in Fig.\ref{fig:1}c). Such a mixing prevents the optical pumping of the $|2\rangle$ sub-level and leads to a considerable scattering of the pump wave light. Therefore, we can see a relatively large amplitude of the EIT resonance in the case of a closed scheme (Fig.\ref{fig:2}a,b). Otherwise, openness of the scheme leads to a small absorption even when $\Omega$$\,\gg\,$$\Delta_c$ because most of the atoms are collected in the $|0\rangle$ trap sub-level by means of optical pumping (Fig.\ref{fig:1}c).

In \cite{Gozzini2009}, the authors proposed to overcome a problem with low contrast of EIT resonance in the standard Hanle scheme by using potassium vapors instead of the cesium ones. That was made possible thanks to a small energy separation between the ground-state hyperfine levels in K, leading to a higher absorption at $B$$\,\ne\,$$0$ than in the case of Cs. In a sense, the system of levels in potassium can be considered as closed. However, a K vapor cell requires either an extended length ($\ge\,$5~cm) or a higher temperature of vapors to achieve a good resonance contrast. Another way is usually realized in miniature SERF magnetometers where very high buffer gas pressure is used (1 atm and more), leading to overlapping of the ground-state hyperfine levels in Cs or Rb atoms \cite{Allred2002,Shah2007}. However, an increased temperature ($\ge\,$150$^\circ$C) is also required to obtain a desirable contrast.

Here, we study a possibility for significant increase in the resonance contrast at lower temperatures ($\le\,$60$^\circ$C) by means of adding the second (probe) light wave. Such ``pump-probe'' light field configurations have been studied in several works, expanding capabilities of the standard Hanle scheme. Most of them utilize two separate light waves \cite{Zibrov2007,Brazhnikov2010,Nicolic2014,Radojicic2015,Gozzini2017,Brazhnikov2018,LeGal2019,Lenci2019,Brazhnikov2019,Brazhnikov2021}. However, it is obvious that a single light wave configuration is much more preferable for developing a miniaturized magnetic field sensor. Our current proposal has many in common with that studied in \cite{Ma2013}: an elliptically polarized light wave can be treated as two co-propagating circularly polarized light waves of opposite handedness. Here, in contrast to \cite{Ma2013}, we propose to use a higher buffer gas pressure, so that the excited-state hyperfine levels in the cesium D$_1$ line are overlapped, while the ground-state hyperfine structure is spectroscopically resolved. As we will see, such a condition appears to be a key point of the proposed technique. Besides, the considered scheme allows either monitoring the LCR in each channel of a polarimeter or using a differential signal of two channels, increasing the SNR.

By analogy with (\ref{eq:3}) and (\ref{eq:4}), the probe wave absorption index reads:

\begin{eqnarray}\label{eq:12}
&&\alpha_p\approx \frac{3\gamma\beta\lambda^2 n_a}{4\pi\gamma_{eg}}\rho_{22}\nonumber\\
&&\approx\frac{\alpha_0\,\bigl[(1+\xi)(1+2\xi)+4\Omega^2\tau^2\bigr]}{(1+\xi)\bigl[1+(2-\beta)\xi\bigr]+4\bigl[1+(1-\beta)\xi\bigr]\Omega^2\tau^2}.
\end{eqnarray}

\noindent This function has the same linewidth as $\alpha_c(\Omega)$ according to expression (\ref{eq:11}). However, since the transmission coefficient $\eta_p$, defined similarly to (\ref{eq:6}), is monitored in the experiments rather than the absorption index $\alpha_p$, the GSHE resonance linewidth can slightly differ from (\ref{eq:11}) because the approximation (\ref{eq:10}) is not valid for the probe wave.

\begin{figure}[!t]
\centering
\includegraphics[width=\linewidth]{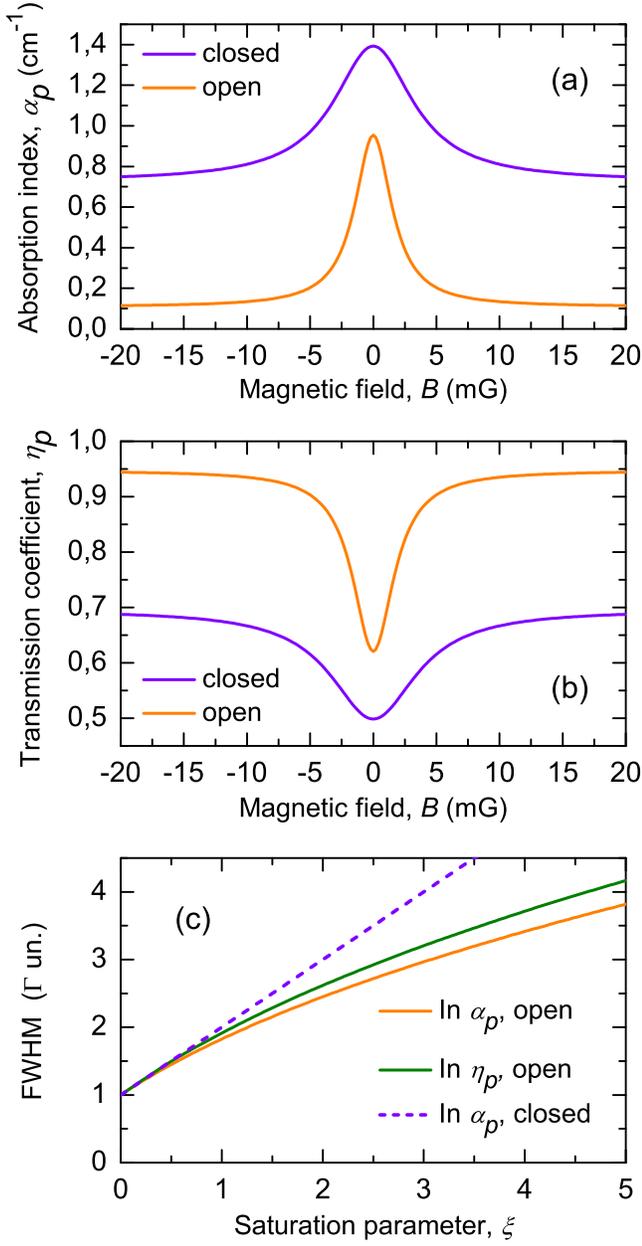}
\caption{\label{fig:3}Calculated zero-field level-crossing resonance of EIA in the probe-wave absorption index (a) and transmission coefficient (b) in closed ($\beta\,$$=$$\,1$) and open ($\beta\,$$=$$\,0.5$) system of energy levels. (c) Full width at half maximum of the resonance as a function of pump wave intensity, namely, the dimensionless saturation parameter $\xi$, observed in the absorption index $\alpha_p$ (orange) and in the transmission coefficient $\eta_p$ (green). Dashed line shows the linewidth behavior ($\propto 1+\xi$) in the closed system of levels. Parameters of the calculation is the same as in Fig.\ref{fig:2}.}
\end{figure}

In contrast to $\alpha_c(\Omega)$, the function $\alpha_p(\Omega)$ exhibits the subnatural-width EIA resonance instead of the EIT one (Fig.\ref{fig:3}a). This behavior has a clear qualitative explanation. For shortness, consider only the case of an open scheme (orange curve in the figure). At $\Omega$$\,=\,$$0$, the probe wave experiences an increased absorption in the cell because many atoms have been prepared in the $|2\rangle$ sub-level by the pump field. Now, if $\Omega$$\,\gg\,$$\Delta_c$, the atoms are mostly transferred to the $|0\rangle$ trap sub-level and the medium becomes almost transparent. This process explains a sign of the resonance as well as its large height in $\alpha_p(\Omega)$.

The transition openness also leads to an increase in the resonance height in the transmission coefficient shown in Fig.\ref{fig:3}b. The behavior of the probe wave absorption, therefore, is very different from that of the pump wave (compare Fig.\ref{fig:2}b and Fig.\ref{fig:3}b). The additional line broadening due to violation of the condition (\ref{eq:10}) for the probe wave is seen in Fig.\ref{fig:3}c at higher pump wave intensities.

Basing on (\ref{eq:7}) and (\ref{eq:12}), we can figure out a ratio between the heights of the resonance in $\alpha_c$ ($A_c$) and $\alpha_p$ ($A_p$):

\begin{equation}\label{eq:13}
    \frac{A_p}{A_c}=1+2(1-\beta)\xi\,.
\end{equation}

\noindent From this, it is clearly seen a constructive action of both the optical pumping process, $\propto\xi$, and the openness of the system, $\propto\,$$(1-$$\beta)$, on the resonance height: in an open system of levels ($\beta$$\ne\,$$1$), the condition $\xi$$\gg$$1$ immediately leads to the relation $A_p$$\,\gg\,$$A_c$, making the proposed pump-probe scheme to be much more attractive than the standard Hanle scheme.

The expression (\ref{eq:7}) for the pump wave absorption index is valid in a wide range of light intensity. At the same time, the expression (\ref{eq:12}) is valid only for such a probe wave intensity that is small enough to do not disturb the atomic sub-level populations. Therefore, it is important to figure out an expression for the light intensity that can characterize a degree of the atom-field interaction strength in the case of an open scheme of levels and a finite time of interaction. In a widely used steady-state two-level model of the atom, such an intensity is known as the saturation intensity:

\begin{equation}\label{Isat2lev}
I_{sat}^{2level}=\frac{4\pi^2\hbar c \gamma_{12}}{3\lambda^3}\,,
\end{equation}

\noindent with $\gamma_{12}$ being the relaxation rate of the optical coherence in the two-level atom that characterizes the homogeneously broadened lineshape of the resonance. At $P$$\,\gg\,$$1$~Torr, it is much higher than the spontaneous relaxation rate $\gamma$ in the case of frequent dephasing collisions of alkali-metal atoms with buffer gas atoms. At $I$$\,=\,$$I_{sat}^{2level}$, a light-field absorption index in the two-level atom drops to a half of its maximum value that takes place at $I$$\ll$$I_{sat}^{2level}$. For the parameters used in our experiments, we can estimate $I_{sat}^{2level}$$\,\approx\,$$340$~mW/cm$^2$ (versus $1.1$~mW/cm$^2$ in a purely spontaneous relaxation regime \cite{Steck}).

Obviously, the expression (\ref{Isat2lev}) is not applicable to the open $\Lambda$-scheme considered here. The valid expression can be obtained in the same way as in the two-level model, namely, we use (\ref{eq:7}) to define saturation intensity as an intensity that satisfies the condition: $\alpha_c(I_{sat})$$\,=\,$$\alpha_0/2$. Trivial calculations lead to the following expression:

\begin{equation}\label{Isat}
I_{sat}=\frac{16\pi^2\hbar c \Gamma \gamma_{eg}}{3\beta\gamma\lambda^3(2-\beta)}\,.
\end{equation}

\noindent This expression can be written in terms of the saturation parameter from (\ref{eq:8}) as $\xi_{sat}$$\,=\,$$(2-\beta)^{-1}$, so that in the closed $\Lambda$-scheme ($\beta$$\,=\,$$1$) we simply get $\xi_{sat}$$\,=\,$$1$. The latter is obvious from (\ref{eq:11}) because at this value the resonance linewidth starts to suffer from the power broadening. For instance, under the experimental conditions used, we can estimate $I_{sat}$ to be around $0.25$~mW/cm$^2$ that is much less than in the two-level model. Note that the developed theory is adequate, only if $I_p$$\,\ll\,$$I_{sat}$. Comparing (\ref{eq:8}) and (\ref{Isat}), we can write an alternative expression for the saturation parameter in our scheme:

\begin{equation}\label{XiIsat}
\xi = \frac{I_c}{(2-\beta)\,I_{sat}}\,.
\end{equation}

At the end of this section, we emphasize once again a positive role of the transition openness in the pump-probe scheme that, on the one hand, shrinks the resonance linewidth and, on the other hand, significantly increases its contrast. It should be noted that it is an unique case because such a transition openness has long been considered as a harmful effect, noticeably degrading properties both the EIT and EIA resonances in various schemes \cite{Papoyan2002,Renzoni1997,Zibrov2005,Auzinsh2012,Kang2017}. In general, the considered line narrowing effect contributes to other known narrowing effects of subnatural-width resonances, for instance, owing to atomic motion in a gas (``Doppler'' narrowing) \cite{Taichenachev2000,Javan2002}, suppression of spin-exchange relaxation \cite{Appelt1999,Allred2002} and influence of the transverse intensity distribution of a light beam \cite{Levi2000,Taichenachev2004}. In particular, in the Doppler narrowing effect, the linewidth is proportional to the square root of light intensity at relatively low intensity values, while it is linearly proportional at higher values. In the case of openness induced narrowing, we have a reverse behavior: the linewidth dependence (\ref{eq:11}) is linear only at $\xi$$\,\ll\,$$1$.

\begin{figure}[!t]
\centering
\includegraphics[width=\linewidth]{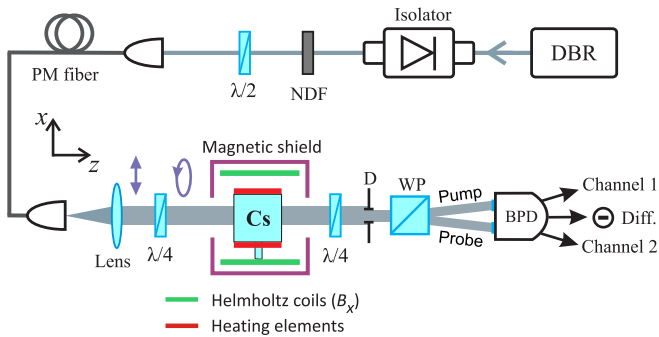}
\caption{\label{fig:4}Experimental setup: DBR, distributed Bragg reflector diode laser; Isolator, optical Faraday insulator; NDF, set of neutral density filters; $\lambda/2$, $\lambda/4$, phase half-wave and quarter-wave plates, respectively; PM fiber, polarization maintaining fiber; Cs, cesium vapor cell; D, iris diaphragm; WP, Wollaston prism; BPD, balanced photodetector.}
\end{figure}

\section{Experiment}\label{Sec:Exp}
\subsection{Experimental setup}\label{Subsec:Setup}

The experimental setup is shown in Fig. \ref{fig:4}. We use a distributed Bragg reflector (DBR) diode laser with a radiation wavelength of $\lambda$$\,\approx\,$$\,894.6$~nm (Cs D$_1$ line) and a linewidth of $\le\,$0.5~MHz. The laser output beam is passed through a Faraday optical isolator. A set of neutral density filters (NDF) is used to control the light power. Subsequently, the beam is sent to a polarization maintaining (PM) optical fiber. A half-wave plate ($\lambda$/2) before the fiber is used to adjust the linear polarization of the beam. A lens is placed after the fiber to collimate the beam (another lens is included into the fiber collimator and can be moved). The beam diameter ($1/e^2$) after the lens is around $1.5$~mm. A quarter-wave ($\lambda$/4) plate placed after the lens creates two circularly polarized light waves, the pump and the probe beams. Relative strength of the beams is determined by the light wave ellipticity and can be controlled by the angle between the main axis of the plate and the direction of linear polarization of the initial light wave. For instance, this angle equals to 45$^\circ$ if one needs a circularly polarized wave as in the standard Hanle scheme. The pump and probe beams pass through the cesium (Cs) vapor cell and their circular polarizations are transformed back into the linear polarizations by the second quarter-wave plate. Since the circularly polarized pump and probe beams have an opposite handedness, their linear polarizations after the second $\lambda$/4 plate are mutually orthogonal. The beams can be then separated in space by means of a Wollaston prism (WP) and monitored independently by using a balanced photodetector (BPD). An iris diaphragm (D) before the WP, having about the same diameter as the beams, is used, on the one hand, to reduce the influence of a Gaussian profile of the beam intensity and, on the other hand, helps to slightly improve the resonance contrast (see details in \cite{Brazhnikov2019}).

\begin{figure}[!b]
\centering
\includegraphics[width=\linewidth]{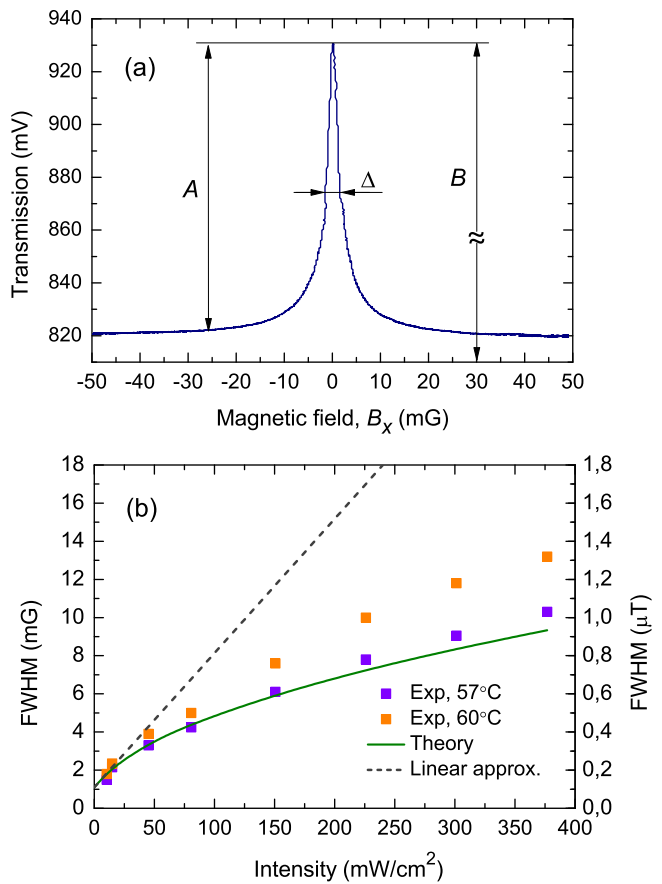}
\caption{\label{fig:5}(a) Level-crossing resonance in the standard Hanle scheme at $P_c$$\,\approx\,$$600$~$\mu$W, $T$$\,\approx\,$$60^\circ$C. (b) Linewidth (FWHM) of the resonance versus the wave intensity. Green line is a square-root fitting of the experimental data points (violet squares) at a lower temperature of 57$^\circ$C. Orange squares stand for the FWHM at an increased temperature. Dashed line is a linear law drawn through the first two experimental points.}
\end{figure}

A cubic $5$$\times$$5$$\times$$5$~mm$^3$ cesium vapor cell is made of ``pyrex'' glass and filled with a neon buffer gas ($\approx\,130$~Torr). The cell is heated by an ac electric current ($100$~kHz) applied to resistive heating elements. The elements are made of a polyimide film, containing the microwires that carry electric currents in the opposite directions to reduce the stray magnetic field. The heating process does not have a visible effect on the LCRs. A three-layer $\mu$-metal magnetic shield is utilized to reduce the ambient field down to $\approx\,0.1$~mG in the cell area.

The absorption profiles corresponding to the separate optical transitions $F_g$$=$$4$$\to$$F_e$$=$$3$ and $F_g$$=$$4$$\to$$F_e$$=$$4$ are significantly broadened due to collisions with a buffer gas, so that they are merged into a single absorption line. The laser frequency is tuned manually to the center of this curve. We use a pair of Helmholtz coils to produce a transverse magnetic field referred here to as $B_x$. It is scanned around zero to observe the GSHE resonance.

Let us provide several estimations to check that our experimental conditions meet the main limitations underlying our theory. At $T$$\,\approx\,$$60^\circ$C, atomic number density in the cell ($n_a$) is around $10^{12}$~cm$^{-3}$ \cite{Steck}. Using the well-known expressions \cite{Happer,VanierBook} and coefficients from \cite{Franz1964}, we get an estimation for the ground-state relaxation rate $\Gamma$ to be around $2\pi$$\times$$420$~Hz (or $\approx1.1$~mG in the magnetic field domain). The D$_1$ line broadening data from \cite{Ha2021} leads to $\gamma_{eg}$$\,\approx\,$$2\pi$$\times$$680$~MHz. In our experiments, the pump wave intensity is in the range $\approx\,$$10$$-$700~mW/cm$^2$. Therefore, from (\ref{eq:8}) one can easily deduce that $R_c$ is in the range $\approx\,$$2\pi$$\times$$(3$$-$24)~MHz $\approx\,$$(1$$-$10)$\gamma$. For the latter estimation we took $\beta$$\,=\,$$0.5$. Finally, in the next subsection we will see that the LCR lays in the region $B_x$$\,\le\,$$50$~mG, i.e. $\Omega$$\,\le\,$$2\pi$$\times$$18$~kHz. The estimations provided demonstrate that the following conditions necessary for validness of our theory are met with a good margin: $\Gamma$$\,\ll\,$$\gamma$$\,\ll\,$$\gamma_{eg}$, $\Omega$$\,\ll\,$$\gamma_{eg}$. Another condition, $R_c^2$$\,\ll\,$$\gamma_{eg}\gamma$, can be rewritten as follows:
$I_c$$\,\ll\,$$\gamma(2-\beta)I_{sat}/\Gamma$$\,\approx\,$$1.5\times10^3$~mW/cm$^2$. Therefore, we can anticipate achieving a good agreement between the theory and the experiments at $I_c$$\,\lesssim\,$$200$~mW/cm$^2$.

\subsection{Openness-induced line-narrowing effect}\label{Subsec:Narrowing}

First, we register a GSHE resonance in the standard Hanle scheme where the light wave has circular polarization. At this regime, the laser beam completely transferred to the channel 1 of the balanced photodetector (see Fig. \ref{fig:4}). At $600$~$\mu$W of optical power, the resonance FWHM is around $3.5$~mG and the contrast is $12\,\%$ (Fig. \ref{fig:5}a). The latter is defined as $C$$\,=\,$$(A/B)$$\times$$100\,\%$ with $A$ being the resonance height and $B$ being the light transmission at center of the resonance. 

Fig. \ref{fig:5}b shows dependence of the resonance linewidth on the laser beam intensity. In the figure, green solid curve fits the experimental data (violet squares) according to the theoretically predicted square-root-like law: $\Delta_c$$\,\approx\,$$\Gamma\,\sqrt{1+I_c/I_0}$. Note that our theory is based on a simplified $\Lambda$ scheme of levels, therefore, we cannot anticipate quantitative agreement between the theory and experiments without using any fitting (free) parameters. Since we have already estimated $\Gamma$ to be around $1.1$~mG, we can use only one fitting parameter $I_0$$\,\approx\,$$5.3$~mW/cm$^2$. As seen from Fig. \ref{fig:5}b, an increase in the cell temperature leads to an additional broadening that looks like a linearization of the linewidth behavior (orange squares) as mentioned in Sec. \ref{Sec:Theory} (see Fig. \ref{fig:3}c). Besides, in the theory section, we have also noted that our expressions are valid if $I_c$ is less than $200$~mW/cm$^2$. Therefore, this can be a reason for appearance of a small deviation from the square-root law in Fig. \ref{fig:5}b when $I_c$ exceeds this value. In general, we can conclude that both sets of the experimental data points in Fig. \ref{fig:5}b are visibly deviate from a linear dependence (dashed line) that could be expected for the closed scheme of energy levels, $\Delta$$\,=\,$$\Gamma\,(1+I_c/I_0)$, i.e. the predicted linewidth narrowing effect is clearly seen from the figure.

\begin{figure}[!t]
\centering
\includegraphics[width=\linewidth]{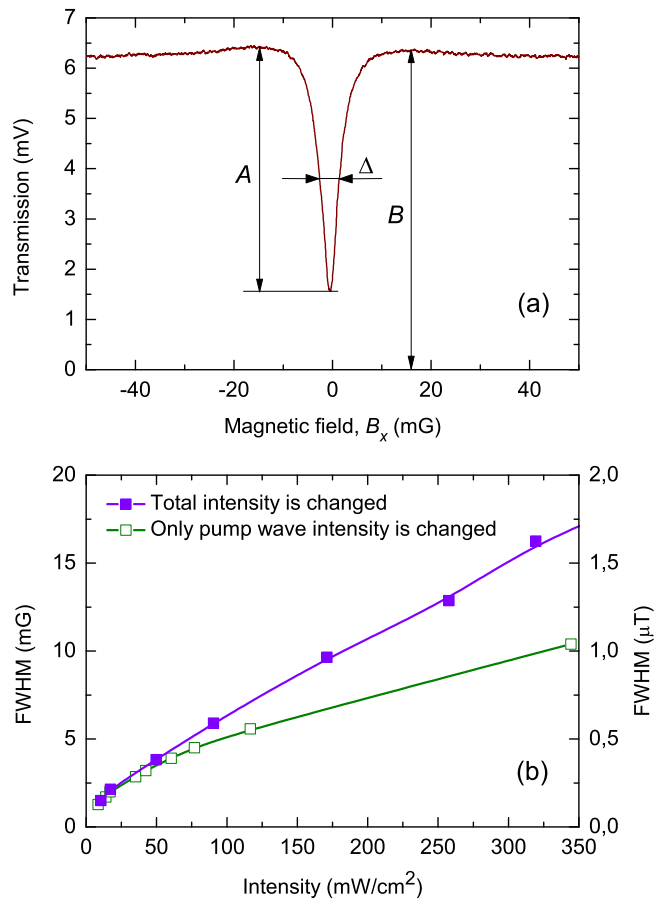}
\caption{\label{fig:6}(a) Level-crossing EIA resonance observed in channel 2 of the PD at $\epsilon$$\,=\,$$39^\circ$, $P_t$$\,=\,$$650$~$\mu$W, $T$$\,=\,$$60^\circ$C. (b) Linewidth (FWHM) of the resonance as the total light intensity is changed (violet) or only the pump wave intensity is changed (green). Solid curves are just guides for a reader's eye.}
\end{figure}

\begin{figure}[!t]
\centering
\includegraphics[width=0.95\linewidth]{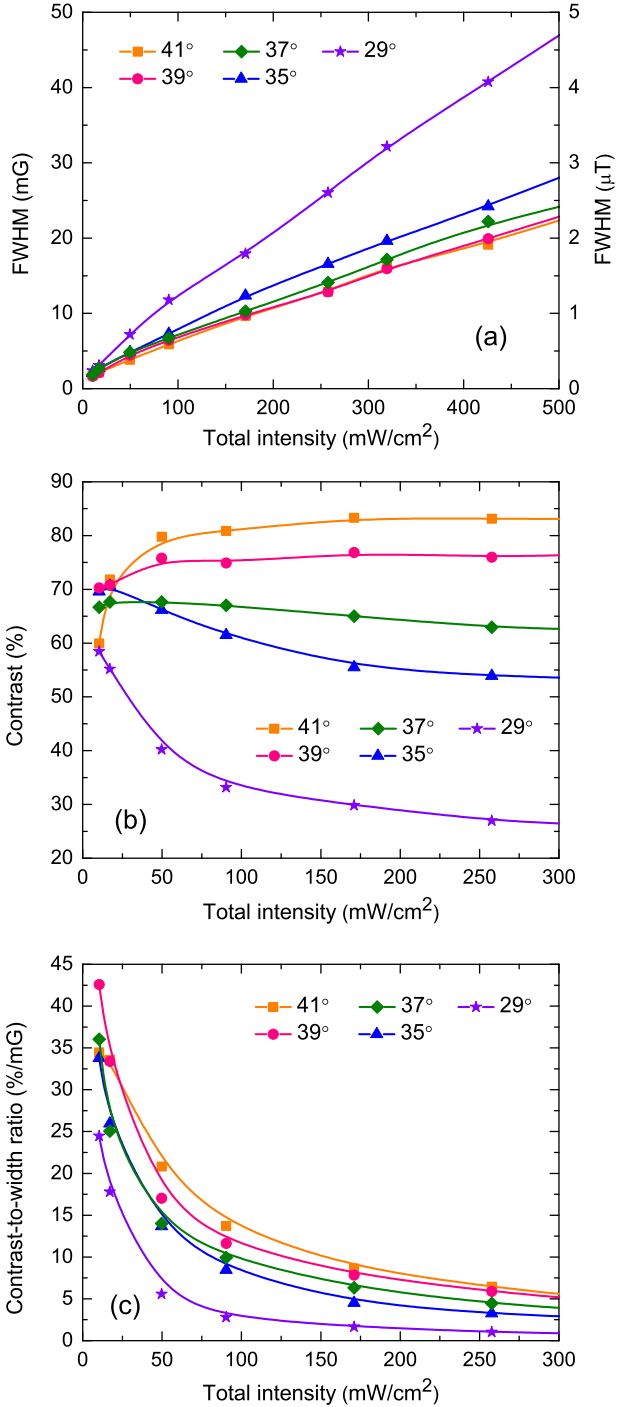}
\caption{\label{fig:7}Parameters of the EIA resonances for different ellipticities versus the total light intensity: Linewidth (a), contrast (b) and contrast-to-width ratio (c). $T$$\,=\,$$60^\circ$C. Solid curves are just guides for a reader's eye.}
\end{figure}

Fig. \ref{fig:6}a shows the level-crossing EIA resonance observed in channel 2 of the photodetector when the polarization of initial beam slightly differs from circular polarization ($\epsilon$$\,\approx\,$$41^\circ$). In the case of EIA resonance, the contrast is defined as $C$$\,=\,$$(A/B)$$\times$$100\,\%$ where $A$ is the resonance height, while $B$ is the background light transmission. As seen from the figure, at $650$~$\mu$W of total optical power, the resonance has $\approx\,80\%$ contrast and $4$~mG linewidth. In contrast to the EIT resonance in the pump wave transmission (at channel 1), the EIA linewidth behavior is almost linear (Fig. \ref{fig:6}b, violet curve). It differs from the linear law only at $I_t$$\,\lesssim\,$$50$~mW/cm$^2$. This peculiarity can be explained by the fact that the probe wave intensity is not as small as it is required by the developed theory. Indeed, it changes as the total intensity ($I_t$) is changed. In particular, the probe wave intensity can be calculated as follows: $I_p$$\,=\,$$I_t\,{\rm sin}^2(\epsilon$$-$$\pi/4)$, so that at $I_t$$\,=\,$$50$~mW/cm$^2$ and $\epsilon$$\,=\,$$39^\circ$, we get $I_p$$\,\approx\,$$0.25$~mW/cm$^2$. This value coincides with the saturation intensity (see Sec. \ref{Sec:Theory}), therefore, the probe wave can be considered to be small in a whole range of $I_t$. Green line in Fig. \ref{fig:6}b corresponds to the case when the probe wave intensity is kept small enough by adjusting the ellipticity of light polarization every time when the total intensity is increased ($I_p$ is in the range $\approx\,$$40$$-$60~$\mu$W/cm$^2$ for the green line). The square-root-like behavior is now clearly manifested as it has been predicted by the theory.

\subsection{Measurement of the resonance parameters}

Here, we focus on analyzing the EIA resonance parameters (linewidth, contrast, contrast-to-width ratio) depending on the total intensity and the ellipticity parameter, keeping the cell temperature around $60^\circ$C. As seen from Fig. \ref{fig:7}a, if the light wave polarization significantly differs from the circular one (blue and violet curves), the linewidth experiences an additional broadening. In the latter case, the probe wave intensity is not small enough and the conditions for observation of the EIA resonance are not optimal. The same reason leads to a degradation of the EIA resonance contrast with the intensity increase in Fig. \ref{fig:7}b. At the same time, at $35^\circ$$<\epsilon<$$45^\circ$ (orange, pink and green curves), the contrast does not experience a visible degradation with the intensity increase, reaching $83\%$ (orange curve).

The EIA resonance in the probe wave transmission can be used for magnetic field measurements. In miniaturized sensors, noise voltage at PD is often higher than the photon-shot-noise limit and contains different contributions linearly proportional to the light intensity. In such a case, a sensitivity of measurements substantially depends on a contrast-to-width ratio (CWR) of the resonance that can be considered as a figure of merit (e.g., see \cite{Lenci2019}). As seen from Fig. \ref{fig:7}c, the EIA resonance CWR takes its maximum value at the lower intensities. It is also seen that the optimal ellipticity is around $40^\circ$.

\begin{figure}[!t]
\centering
\includegraphics[width=0.95\linewidth]{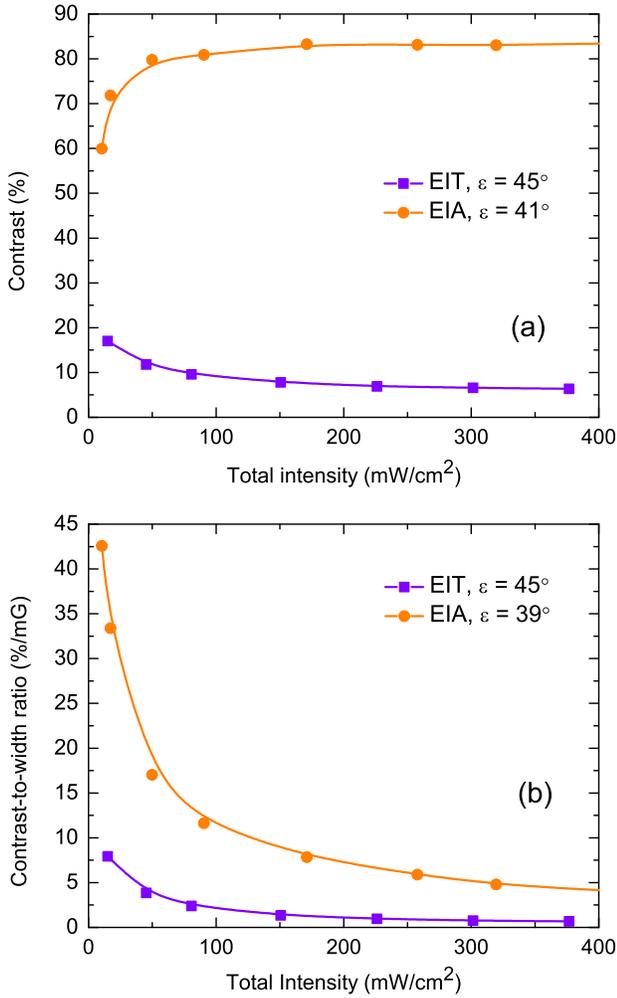}
\caption{\label{fig:8}Comparison of the EIT resonance parameters (in the standard Hanle scheme with a circularly polarized wave) with the best EIA parameters observed in the experiments. $T$$\,=\,$$60^\circ$C. Solid curves are just guides for a reader's eye.}
\end{figure}

\begin{figure}[!t]
\centering
\includegraphics[width=0.95\linewidth]{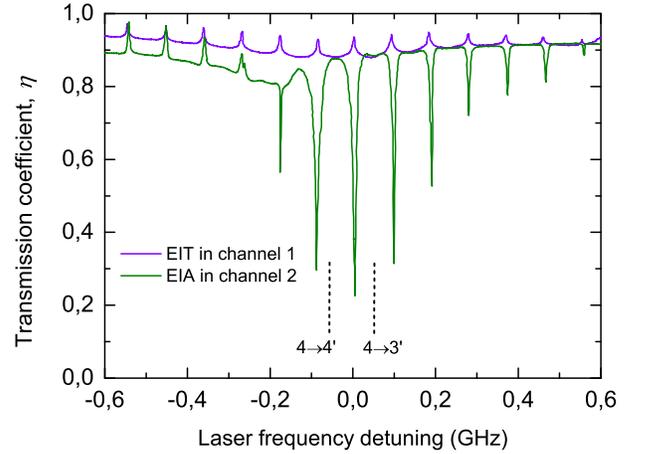}
\caption{\label{fig:9}Transmission signals observed at $\epsilon$$\,=\,$$37^\circ$ in both channels of the balanced PD with simultaneous scanning the light field frequency detuning ($f_{scan}$$\,=\,$$0.5$~Hz) and the transverse magnetic field ($f_{scan}$$\,=\,$$20$~Hz). The separate optical transitions $F_g$$\,=\,$$4$$\,\to\,$$F_e$$\,=\,$$3$ and $F_g$$\,=\,$$4$$\,\to\,$$F_e$$\,=\,$$4$ are not resolved due to buffer-gas line broadening. $P_t$$\,=\,$$5.3$~mW, $T$$\,=\,$$63^\circ$C.}
\end{figure}

Let us compare C and CWR parameters measured for the EIT (channel 1) and the EIA (channel 2) resonances at the same temperature of the cell. As follows from Fig. \ref{fig:8}, the contrast and the CWR are much higher in the case of EIA. High contrast of the EIA resonances in comparison with the EIT ones is also demonstrated in Fig. \ref{fig:9} where the laser frequency is scanned simultaneously with scanning the magnetic field. If $B_x$$\,\approx\,$$0$, then the pump wave experiences an increased transmission (EIT), while transmission of the probe wave is dramatically decreases (EIA). One can easily compare heights of the EIT and EIA resonances. It can be noted that a non-zero optical frequency detuning from the center of absorption line may cause change in the resonance sign. In such a case only one optical transition, $F_g$$\,=\,$$4$$\,\to\,$$F_e$$\,=\,$$3$ or $F_g$$\,=\,$$4$$\,\to\,$$F_e$$\,=\,$$4$, is predominantly excited by the light, exhibiting EIT or EIA type resonances in dependence of the angular momentum values of the involved levels. Besides, there can be a non-negligible influence of the circular birefringence under a non-zero frequency detuning, competing with the dichroism effects.

\begin{figure}[!t]
\centering
\includegraphics[width=0.95\linewidth]{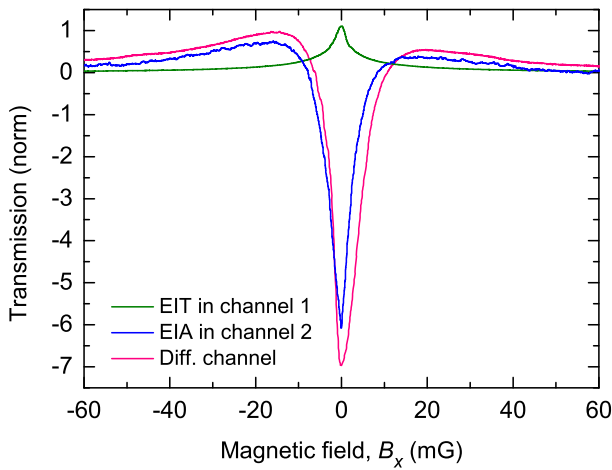}
\caption{\label{fig:10}Level-crossing resonances observed in different channels of the balanced photodetector at $\epsilon$$\,=\,$$37^\circ$, $P_t$$\,=\,$$1.1$~mW, $T$$\,=\,$$60^\circ$C.}
\end{figure}

In the end of this subsection, we would like to mention another way to observe the GSHE resonance which consists in monitoring a differential channel at PD. Fig. \ref{fig:10} shows the normalized signals from three channels. It is seen that the differential channel provides the resonance with an increased height that is nothing but the sum of the EIT and EIA resonance heights. This observation technique deserves an additional study and is not considered in the current paper.

\subsection{Sensitivity}

\begin{figure}[!t]
\centering
\includegraphics[width=\linewidth]{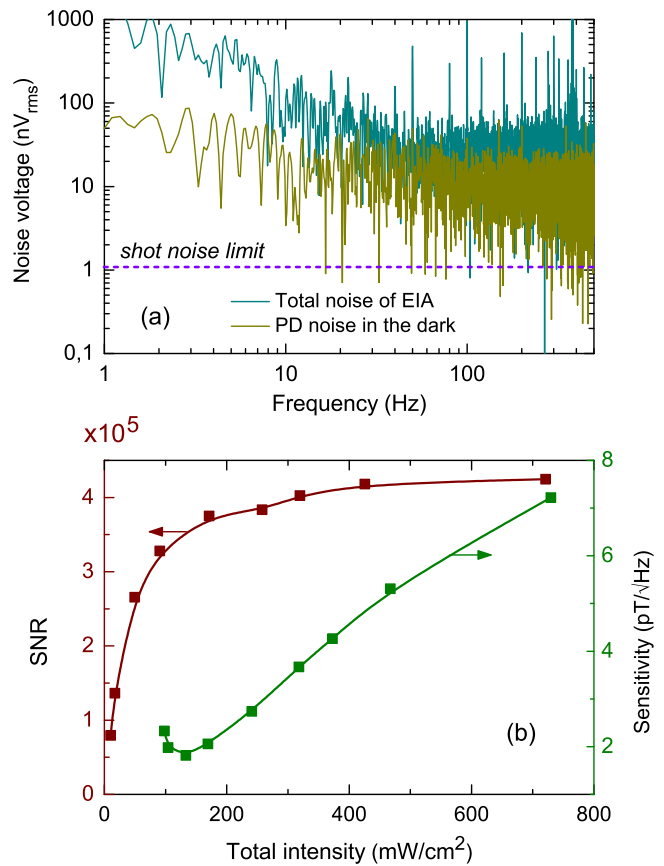}
\caption{\label{fig:11}(a) Noise voltage of the EIA resonance (dark cyan) at $P_t$$\,=\,$$1$~mW and the dark noise of the PD (dark yellow). (b) SNR (dark red) and sensitivity (green) at 40 Hz as a work frequency of the sensor versus total intensity of the beam. Conditions of measurements: $\epsilon$$\,=\,$$37^\circ$, $T$$\,=\,$$60^\circ$C.}
\end{figure}

To estimate the sensitivity, we have measured the noise voltage and signal-to-noise ratio in channel 2 at $\epsilon$$\,=\,$$37^\circ$ (Fig. \ref{fig:11}a,b). On a log-log plot in Fig. \ref{fig:11}a, the noise decreases linearly down to $\approx\,30$~nV$_{\rm rms}$ at $40$~Hz. It has been checked in our experiments (not shown here) that the the EIA resonance characteristics do not degrade visibly up to $150$~Hz of the scanning frequency. Therefore, we can take a work frequency of the sensor in the range $40-150$~Hz. Then, using the linewidth dependence in Fig. \ref{fig:7}a, we can now plot the sensitivity, $\delta B$, versus the light field intensity (Fig. \ref{fig:11}c). It reaches $\approx\,1.8$~pT/$\surd$Hz at $I_t$$\,\approx\,$$650$~$\mu$W. This sensitivity is determined mainly by the laser intensity noise, dark noise of the PD, and electric current noise in the Helmholtz coils which is transferred to the intensity noise at the resonance slope. As an example, we show the dark noise of the PD in the same plot. A shot noise limit showed in the figure as a dashed line is much less than the observed total noise. Therefore, if all necessary steps are taken to reduce the noise sources, we can expect to achieve a shot-noise-limited sensitivity of $\approx\,60$~fT/$\surd$Hz. Some improvement likely can be obtained just by using a differential channel of the PD. However, we do not focus on this possibility in the current study because it deserves separate and careful investigations.

%measured by a filament lamp is around $1$~nV$_{\rm rms}$

\section{Conclusions}

In this paper, we have considered subnatural-width zero-field level-crossing resonances (LCRs) in a cesium vapor cell with a buffer gas. The effect is also known as the ground-state Hanle effect. The atoms are excited by a single light wave, while a transverse magnetic field is slowly scanned around zero to observe the LCR in the light beam transmission. The buffer gas pressure is such that the excited-state hyperfine levels of the Cs D$_1$ line are overlapped due to collisional broadening, while the ground-state hyperfine levels are spectroscopically resolved. Basing on a three-level ($\Lambda$) scheme, we have developed a simplified theory that provides us with an explicit analytical solution for the light-wave absorption index. The solution revealed a line narrowing effect that can be observed in an open system of energy levels. The effect has then been verified experimentally.

The main goal of the paper consisted in developing a single-beam technique that could provide high-quality LCRs in a low-temperature small vapor cell. Such a technique can be of a certain interest from the side of boimedical applications where an array of magnetic field sensors are required. We have proposed to use a single elliptically polarized light wave that, under the conditions considered, can be treated as a combination of two independent circularly polarized waves: the pump wave creates a strong circular dichroism in the resonant medium, while the probe wave is to interact with the prepared atoms. It has been shown that LCRs in the probe wave transmission demonstrate an extraordinary contrast up to $83\%$ versus $17\%$ as in the standard Hanle scheme. The observed contrast-to-width ratio, a figure of merit of the resonance, reached $42$~\%/mG. It should be emphasized that the observed high characteristics of the resonances have been obtained in a small $5\times5\times5$~mm$^3$ vapor cell heated to a relatively low temperature of $\approx\,60^\circ$C, in contrast to many other schemes where alkali-metal vapors are usually heated to a temperature higher than $120^\circ$C to reach the SERF regime of operation. The absence of the SERF regime in our case also means a potentially large dynamic range of measurements of the sensor up to several $\mu$T.

The proposed technique for observing the LC resonances can be used for developing a high-sensitivity compact magnetic-field sensor of low power consumption and heat release. To demonstrate this, we have measured the noise voltage observed at a photodetector. The noise level appeared to be significantly higher than the shot-noise limit, nevertheless, the estimated sensitivity reached $\approx\,1.8$~pT/$\surd$Hz at $40$~Hz. We expect that additional efforts directed to the noise suppression will help us to demonstrate the shot-noise-limited sensitivity of $\approx\,60$~fT/$\surd$Hz. However, such a systematic work deserves a separate study.

In our experiments, a pair of Helmholtz coils are used to scan the transverse field ($B_x$) for observing the resonance. The resonance shift provides information about the $x$-component of ambient magnetic field that can be measured. A 2D operation of the sensor can be easily realized by using the second pair of coils for scanning $B_y$ field. Full 3D (vector) mode of operation can be also realized by different ways. One of them is to use additional coils to compensate the $z$-component of ambient field by maximization of the resonance contrast. Another way is to monitor the resonance linewidth that experiences additional linear broadening if $B_z$ is presented in the cell \cite{Brazhnikov2021}. Also, all three components of the field can be measured by scanning the transverse and longitudinal magnetic fields to observe different types of the LCRs \cite{LeGal2022}.

In the end, we would like to note that a small vapor cell used in our experiments can be integrated into a compact magnetic field sensor as it has been already demonstrated by other groups with similar cubic glass vapor cells (e.g., see \cite{Shah2009,Boto2017}). However, it is truly fascinating to be able to use the proposed technique in a combination with booming photonic-atomic chip-scale technologies that provide an extreme miniaturization. For instance, a photon spin ``sorter'' has been recently demonstrated in \cite{Sebbag2021}. Such a device is probably best suited for creating a chip-scale circular-dichroism-based quantum sensor.

\begin{acknowledgments}
The work has been supported by Russian Science Foundation (17-72-20089). We also thank Russian Foundation for Basic Research (20-52-18004) and Bulgarian National Science Fund (KP-06-Russia/11).
%A.N.G. thanks Ministry of Science and Higher Education of the Russian Federation (Project No. AAAA-A19-119102890006-5).
\end{acknowledgments}

\appendix*
%\section{Density Matrix Formalism}
\section{}

\renewcommand{\theequation}{A\arabic{equation}}

Here, we apply a well-known density matrix approach \cite{Blum} to figure out analytical expressions for absorption indices of the pump and probe waves in atomic vapors. We start from the general equation (\ref{eq:2}), where the free-atom Hamiltonian has the following explicit form:

\begin{equation}\label{H0}
\hat{H}_0=\sum\limits_{n=0}^3{\cal E}_n|n\rangle \langle n|\,,
\end{equation}

\noindent where ${\cal E}_n$ is an energy of $n$ level with ${\cal E}_1$$\,=\,$${\cal E}_2$ (see Fig. \ref{fig:1}b), so that $\omega_{31}$$\,=\,$$\omega_{32}$$\,=\,$$\bigl({\cal E}_3$$-$${\cal E}_1\bigr)/\hbar$ being the optical transition frequencies in the $\Lambda$ scheme. Angular brackets $\langle \dots |$ and $|\dots\rangle$ stand for the Dirac's bra and ket vectors, respectively.

The operator $\hat{V}_b$ is responsible for the interaction between the atomic spins and ambient transverse magnetic field (${\bf B}_x$$\perp$${\bf k}$). It leads to mixing the $|1\rangle$ and $|2\rangle$ states: 

\begin{equation}\label{Vb}
\hat{V}_b=\hbar\Omega\Bigl( |1\rangle \langle 2| + |2\rangle \langle 1| \Bigr)\,,
\end{equation}

\noindent where $\Omega$ is the Larmor frequency.

In the rotating wave approximation (RWA), the interaction between the atomic dipole momentum ($d$) and the light field (\ref{eq:1}) is described by the operator:

\begin{eqnarray}\label{Ve}
\hat{V}_e = -\hat{d}E(t,z) &=& -\hbar R_c e^{-i\omega t} |3\rangle \langle 1| \nonumber\\
&-& \hbar R_p e^{-i\omega t} |3\rangle \langle 2| + H.c.\,,
\end{eqnarray}

\noindent with $R_{c,p}$ being the Rabi frequencies for the pump ($c$) and probe ($p$) waves, ``$H.c.$'' means the Hermitian conjugate terms. As shown in Sec. \ref{Sec:Theory}, the pump wave experiences small absorption in the cell. The probe wave in turn is weak enough to do not affect the atomic density matrix at all. Therefore, we do not take into account the dependence of the Rabi frequencies on the $z$ coordinate in (\ref{Ve}).

The density matrix $\hat{\rho}$ for the considered $\Lambda$ scheme can be written in the following form:

\begin{equation}\label{eq:rho}
\hat{\rho}=\left(
\begin{array}{cccc}
\rho_{00} & 0 & 0 & 0 \\
0 & \rho_{11} & \rho_{12} & \rho_{13} \\
0 & \rho_{21} & \rho_{22} & \rho_{23} \\
0 & \rho_{31} & \rho_{32} & \rho_{33}
\end{array}\right)\,,
\end{equation}

\noindent where diagonal elements $\rho_{nn}$ ($n$$\,=\,$$0,\dots,3$) denote the sub-level populations, $\rho_{12}$ and $\rho_{21}$ stand for the so-called Zeeman coherences, $\rho_{13}$, $\rho_{31}$, $\rho_{23}$ and $\rho_{32}$ are known as optical coherences because they oscillate in time at the optical frequency $\omega$. Other non-diagonal elements, such as $\rho_{0n}$ and $\rho_{n0}$ ($n$$\,=\,$$1,\dots,3$), equal to zero because there is no any light or microwave field in the system that could couple the corresponding sub-levels. Therefore, $|0\rangle$ can be referred to as the ``trap'' state with the population $\rho_{00}$. Since the density matrix is Hermitian, $\hat{\rho}$$\,=\,$$\hat{\rho}^\dag$, we get the following relations: $\rho_{nm}$$\,=\,$$\rho_{mn}^*$.

The relaxation operator $\hat{\cal R}$ in (\ref{eq:2}) reflects influence of three different processes in the atom: spontaneous radiation emission from the excited state occurring at the rate $2\gamma$, collisional broadening of the optical absorption line ($\gamma_c$), and diffusive motion of alkali-metal atoms in buffer gas ($\Gamma$), during which a spin-polarized atom either leaves the light beam or undergoes a spin-exchange/destruction collision (see review \cite{Happer}). We can consider these contributions by separate terms, namely, the spontaneous relaxation reads:

\begin{equation}\label{eq:sponrelax}
\hat{\cal R}_{spon}=\gamma\,\left(
\begin{array}{cccc}
\beta_0\rho_{33} & 0 & 0 & -1 \\
0 & \beta_1\rho_{33} & 0 & -1 \\
0 & 0 & \beta_2\rho_{33} & -1 \\
-1 & -1 & -1 & -2
\end{array}\right)\,,
\end{equation}

\noindent where the branching ratios are $\beta_1$$\,=\,$$\beta_2$$\,\equiv\,$$\beta$, $\beta_0$$\,=\,$$2(1-\beta)$.

The collisional line broadening is described by the following terms:

\begin{equation}\label{eq:collrelax}
\hat{\cal R}_{coll}=-\gamma_c\sum_{n=1}^2\rho_{n3}|n\rangle\langle 3|+H.c.\,,
\end{equation}

\noindent with ``H.c.'' meaning the Hermitian conjugate terms.

Finally, the diffusion time-of-flight relaxation has the form:

\begin{equation}\label{eq:diffusrelax}
\hat{\cal R}_{diff}=-\Gamma\hat{\rho}+\frac{1}{3}\Gamma\sum_{n=0}^2|n\rangle\langle n|\,.
\end{equation}

\noindent Here, we take into account that the initial (isotropic) population distribution is so that $\rho_{00}$$\,=\,$$\rho_{11}$$\,=\,$$\rho_{22}$$\,=\,$$1/3$.

We apply the rotating wave approximation (RWA) that means the following series expansion for the optical coherences:

\begin{eqnarray}\label{eq:RWA}
&&\rho_{13}(t)=\tilde{\rho}_{13}\,e^{i\omega t},\quad \rho_{23}(t)=\tilde{\rho}_{23}\,e^{i\omega t}\nonumber\\
&&\rho_{31}(t)=\tilde{\rho}_{31}\,e^{-i\omega t},\quad \rho_{32}(t)=\tilde{\rho}_{32}\,e^{-i\omega t}.
\end{eqnarray}

We consider the atom-field interaction under the optical resonance condition when $\omega$$\,=\,$$\omega_{31}$$\,=\,$$\omega_{32}$. Using the master equation (\ref{eq:2}) and all the above-mentioned expressions, one may come to the following set of linear differential equations:

\begin{eqnarray}\label{eq:system}
&&\frac{d}{dt}\rho_{00}=\frac{1}{3}\Gamma+2(1-\beta)\gamma\rho_{33}-\Gamma\rho_{00}\,,\\
&&\frac{d}{dt}\rho_{11}=\frac{1}{3}\Gamma+\beta\gamma\rho_{33}-\Gamma\rho_{11}\nonumber\\
&&\qquad\qquad+i\Omega(\rho_{12}-\rho_{21})+iR_c\bigl(\tilde{\rho}_{31}-\tilde{\rho}_{13}\bigr)\,,\\
&&\frac{d}{dt}\rho_{22}=\frac{1}{3}\Gamma+\beta\gamma\rho_{33}-\Gamma\rho_{22}\nonumber\\
&&\qquad\qquad+i\Omega(\rho_{21}-\rho_{12})+iR_p\bigl(\tilde{\rho}_{32}-\tilde{\rho}_{23}\bigr)\,,\\
&&\frac{d}{dt}\rho_{33}=-2\gamma\rho_{33}-\Gamma\rho_{33}+iR_c\bigl(\tilde{\rho}_{13}-\tilde{\rho}_{31}\bigr)\nonumber\\
&&\qquad\qquad\qquad\qquad+iR_p\bigl(\tilde{\rho}_{23}-\tilde{\rho}_{32}\bigr)\,,\\
&&\frac{d}{dt}\rho_{12}=-\Gamma\rho_{12}+ i\Omega(\rho_{11}-\rho_{22})\nonumber\\ &&\qquad\qquad\qquad\qquad+iR_c\tilde{\rho}_{32}- iR_p\tilde{\rho}_{13}\,,\\
&&\frac{d}{dt}\tilde{\rho}_{13}=-\gamma_{eg}\tilde{\rho}_{13}-i\Omega\tilde{\rho}_{23}\nonumber\\ &&\qquad\qquad\qquad+iR_c(\rho_{33}-\rho_{11})- iR_p\rho_{12}\,,\\
&&\frac{d}{dt}\tilde{\rho}_{23}=-\gamma_{eg}\tilde{\rho}_{23}-i\Omega\tilde{\rho}_{13}\nonumber\\ &&\qquad\qquad\qquad+iR_p(\rho_{33}-\rho_{22})- iR_c\rho_{21}\,.
\end{eqnarray}

\noindent Other three equations on $\tilde{\rho}_{31}$, $\tilde{\rho}_{32}$ and $\tilde{\rho}_{21}$ can be obtained by complex conjugation of corresponding equations.

The pump wave propagates through the medium according to the reduced wave equation:

\begin{equation}\label{eq:ReducedEq}
    \frac{dE_c}{dz}=2\pi i k \tilde{P}_c\,,
\end{equation}

\noindent where $P_c(z,t)$$\,=\,$$\tilde{P}_c(z)\,e^{-i\omega t}$ is the medium polarization induced by the pump wave with $\tilde{P}_c$$\,=\,$$n_a d_0 \tilde{\rho}_{31}$. Taking into account the relation $I_{p,c}$$\,=\,$$(c/2\pi)E_{p,c}^2$, we get the equation for the pump wave intensity change

\begin{equation}\label{eq:ReducedEq2}
    \frac{dI_c}{dz}=2n_a\hbar\omega R_c \rm{Im}\bigl\{\tilde{\rho}_{13}\bigr\}\,.
\end{equation}

Considering the steady-state regime, when the complex amplitudes $\tilde{\rho}_{nm}$ in (\ref{eq:RWA}) do not depend on time as well as the sub-level populations ($\rho_{nn}$) and Zeeman coherences ($\rho_{12}$, $\rho_{21}$), we come to the expression

\begin{eqnarray}\label{eq:ro13}
   \tilde{\rho}_{13}&&=\frac{i\gamma_{eg}R_c}{\Omega^2+\gamma_{eg}^2}\bigl(\rho_{33}-\rho_{11}\bigr)-\frac{R_c\Omega}{\Omega^2+\gamma_{eg}^2}\rho_{21}\nonumber\\
   &&+\frac{R_p\Omega}{\Omega^2+\gamma_{eg}^2}\bigl(\rho_{33}-\rho_{22}\bigr)-\frac{i\gamma_{eg}R_p}{\Omega^2+\gamma_{eg}^2}\rho_{12}\,.
\end{eqnarray}

In a linear approximation on the probe field, we can take $R_p$$\,=\,$$0$ in the latter formula. We also assume that $\Omega$$\,\ll\,$$\gamma_{eg}$ and $R_c$$\,\ll\,$$\gamma_{eg}$, so that (\ref{eq:ro13}) now reads

\begin{equation}\label{eq:ro13approx}
   \tilde{\rho}_{13}\approx i\frac{R_c}{\gamma_{eg}}\bigl(\rho_{33}-\rho_{11}\bigr)\,.
\end{equation}

\noindent This expression means that the optical coherence on the transition $|1\rangle$$\to$$|3\rangle$ is created by the pump light field rather than the interference action of the probe light field and the transverse magnetic field. In our approach, this coherence contains only the imaginary part, i.e. any effects of birefringence, like the Voigt (Cotton-Mouton) effect \cite{Budker2002}, are negligible in comparison to dichroism effects under the optical resonance condition, $\omega$$\,=\,$$\omega_{32}$$\,=\,$$\omega_{31}$, considered in the present work.

By substituting (\ref{eq:ro13approx}) into (A12) and taking $R_p$$\,=\,$$0$, in the steady state, we get

\begin{equation}\label{eq:ro33linear}
  \rho_{33}\approx \frac{2R_c^2\tau}{\gamma_{eg}(1+2\gamma\tau)}\bigl(\rho_{11}-\rho_{33}\bigr)\,.
\end{equation}

\noindent In this equation we can assume $\gamma\tau$$\gg$$1$ that is satisfied with a good margin in the case of a buffered vapor cell. Furthermore, due to a large collisional broadening of the optical absorption line, we also have the condition $R_c^2$$\ll$$\gamma\gamma_{eg}$. These assumptions lead to a simple solution

\begin{equation}\label{eq:ro33linear2}
  \rho_{33}\approx \frac{R_c^2}{\gamma\gamma_{eg}}\rho_{11}\,,
\end{equation}

\noindent meaning that $\rho_{33}$$\ll$$\rho_{11}$. Therefore, we can neglect the influence of the excited-state population on the pump wave absorption in the cell. All these conclusions result in the following law:

\begin{equation}\label{eq:ReducedEq3}
    \frac{dI_c}{dz}\approx-\frac{2n_a\hbar\omega}{\gamma_{eg}}\rho_{11}R_c^2=-\frac{3\gamma\beta\lambda^2 n_a}{4\pi\gamma_{eg}}\rho_{11}I_c\,,
\end{equation}

\noindent which coincides with (\ref{eq:3}).

To find the steady-state population $\rho_{11}$, we assume that the Zeeman coherences $\rho_{12}$ and $\rho_{21}$ are produced only by the magnetic field. This assumption is valid because, as mentioned in the beginning of Sec.\ref{Sec:Theory}, the interference between the pump and the probe waves can be neglected. If $R_c$$\ll$$\gamma_{eg}$ and $\Omega$$\ll$$\gamma_{eg}$, then we find

\begin{eqnarray}\label{eq:ro12solution}
\rho_{12}\approx\frac{i\Omega\tau}{1+\xi}\bigl(\rho_{11}-\rho_{22}\bigr)\,,
\end{eqnarray}

\noindent with $\rho_{21}$$\,=\,$$\rho_{12}^*$.

Substituting (\ref{eq:ro13approx}), (\ref{eq:ro12solution}) into (A10) and assuming $\rho_{33}$$\ll$$\rho_{11}$ we come to the solution

\begin{equation}\label{eq:ro11solution}
\rho_{11}\approx \rho_0\,\frac{1+\xi+4\Omega^2\tau^2}{(1+\xi)\chi+4\Omega^2\tau^2(\chi-\xi)}\,
\end{equation}

\noindent with $\chi$$\,=\,$$\bigl[1+(2-\beta)\xi\bigr]$. In combination with (\ref{eq:ReducedEq3}), the latter expression results in the pump-wave absorption index (\ref{eq:7}). In (\ref{eq:ro11solution}), $\rho_0$$\,=\,$$1/3$ is the initial sub-level population in absence of the light field.

Similarly to (\ref{eq:ReducedEq2}), the probe wave absorption satisfies the following equation: 

\begin{equation}\label{eq:ReducedProbe}
    \frac{dI_p}{dz}=2n_a\hbar\omega R_p \rm{Im}\bigl\{\tilde{\rho}_{23}\bigr\}\,.
\end{equation}

\noindent At $\Omega$$\ll$$\gamma_{eg}$, from (A15) we get

\begin{equation}\label{eq:ro23approx}
   \tilde{\rho}_{23}\approx i\frac{R_p}{\gamma_{eg}}\bigl(\rho_{33}-\rho_{22}\bigr)+i\frac{R_c}{\gamma_{eg}}\rho_{12}\,,
\end{equation}

\noindent and (\ref{eq:ReducedProbe}) now transforms to

\begin{equation}\label{eq:ReducedProbe2}
    \frac{dI_p}{dz}=\frac{2n_a\hbar\omega}{\gamma_{eg}} \Bigl[R_p^2\bigl(\rho_{33}-\rho_{22}\bigr)- R_pR_c\,\rm{Re}\bigl\{\rho_{12}\bigr\}\Bigr]\,.
\end{equation}

\noindent Neglecting small impact from the excited-state population and influence of the interference term ($\propto R_cR_p$) gives

\begin{equation}\label{eq:ReducedProbe3}
    \frac{dI_p}{dz}\approx-\frac{2n_a\hbar\omega}{\gamma_{eg}}\rho_{22}R_p^2=-\frac{3\gamma\beta\lambda^2 n_a}{4\pi\gamma_{eg}}\rho_{22}I_p\,,
\end{equation}

\noindent where $\rho_{22}$ does not depend on the probe field in the linear regime. Besides, since the pump field experiences small absorption in the cell (see discussions in the main text), $\rho_{22}$ does not depend on $z$ as well. Therefore, one can easily get an explicit solution in the form

\begin{equation}
    I_p(z)=I_{p0}\,e^{-\alpha_p z}\,,
\end{equation}

\noindent with $I_{p0}$ being the probe wave intensity at entrance of the cell and $\alpha_p$ being the probe-wave absorption index: 

\begin{equation}\label{eq:probeindex}
\alpha_p\approx \frac{3\gamma\beta\lambda^2 n_a}{4\pi\gamma_{eg}}\rho_{22}.
\end{equation}

Using the same assumptions ($R_c^2$$\ll$$\gamma\gamma_{eg}$, $\Omega$$\ll$$\gamma_{eg}$), the steady-state solution of (A11) gives

\begin{equation}\label{eq:ro22solution}
\rho_{22}\approx \rho_0\,\frac{(1+\xi)(1+2\xi)+4\Omega^2\tau^2}{(1+\xi)\chi+4\Omega^2\tau^2(\chi-\xi)}\,,
\end{equation}

\noindent which results in the probe-wave absorption index (\ref{eq:12}).

\end{document}